\documentclass[10pt]{article}
\usepackage{graphicx}
\begin{document}

\title{\bf A Dynamical Systems Approach to an Inviscid and Thin
Accretion Disc}

\author{Arnab K. Ray\thanks{tpakr@mahendra.iacs.res.in} 
\ and J. K. Bhattacharjee\thanks{tpjkb@mahendra.iacs.res.in}\\
Department of Theoretical Physics\\
Indian Association for the Cultivation of Science\\
Jadavpur, Calcutta (Kolkata) 700032, INDIA\\}

\date{}
\maketitle

\begin{abstract}
The inviscid and thin accretion disc is a simple and well
understood model system in accretion studies. In this work,
modelling such a disc like a dynamical system, we analyse the
nature of the fixed points of the stationary solutions of the flow.
We show that of the two fixed points, one is a saddle and the
other is a centre type point. We then demonstrate, using a
simple but analogous mathematical model, that a temporal
evolution of the flow is a very likely non-perturbative mechanism
for the selection of an inflow solution that passes through the
saddle type critical point.
\end{abstract}


\section{Introduction}

In astrophysical fluid dynamics, studies in accretion processes
occupy a position of prominence. Of such processes, critical
(transonic) flows --- flows which are regular through a critical
point --- are of great importance~\cite{skc}. A classic example of 
such a flow, in the stationary spherically symmetric situation, 
is the Bondi flow~\cite{bon}. What is striking about the Bondi solution
is that while it is virtually impossible to numerically generate it
in the stationary limit of the hydrodynamic flow, it is easy to lock 
on to the Bondi solution when the temporal evolution of the flow is
followed. This dynamic and non-perturbative selection 
mechanism of the transonic solution is in very favourable agreement 
with Bondi's conjecture that it is the criterion of minimum energy that 
will make the transonic solution the favoured one~\cite{bon,rb}. 

The appeal of the spherically symmetric flow, however, is somewhat
pedagogic, limited as it is by its not accounting for  
the fact that in a reasonably realistic
situation, the infalling matter would be in possession of angular
momentum --- and hence the process of infall should lead to the formation
of what is known as an accretion disc. A well understood and quite
regularly invoked model of such a system is the inviscid and thin accretion
disc~\cite{skc,az,msc,skc2}. 
In this work, we show how such a physical system --- studied in its
stationary limit --- gives rise to saddle point like behaviour for at least
one of the critical points of the flow. The adverse implications regarding
the realizability of a solution passing through such a point is then 
addressed by considering the system in a real non-linear time evolution.
A simplified mathematical analog of such a situation shows that the 
evolution indeed selects the critical (transonic) solution. 
Finally, we discuss that in the vicinity of the critical point, the 
resulting perturbation equation arising
out of a linear stability analysis of the stationary solutions, 
quite closely resembles the metric of an
acoustic black hole, and that is probably the signal that trajectories
pass through the critical point in a unidirectional fashion. 

\section{The stationary equations of the flow and its critical points}

The flow system to be considered is an axisymmetric thin disc of a 
compressible astrophysical fluid, with a thickness $H$~\cite{fkr}.
In the vertical 
direction the flow is considered to be in hydrostatic equilibrium. 
The governing equations for the flow should therefore be the equation
of continuity and the equations for momentum balance in both the 
radial and the azimuthal directions. It is customary to write the 
stationary flow equations in the following notation~\cite{ny}.
\begin{itemize}

\item Equation of continuity:
\begin{equation}
\frac{d}{dR}(\rho RHv)=0
\end{equation}

\item Radial momentum balance equation:
\begin{equation}
v{\frac{dv}{dR}}+{\frac{1}{\rho}}{\frac{dP}{dR}} 
+ \left({{{\Omega}_K}^2} -{{\Omega}^2} \right)R = 0 
\end{equation}

\item Angular momentum balance equation:
\begin{equation}
v{\frac{d}{dR}}(\Omega R^2)={\frac{1}{\rho RH}}{\frac{d}{dR}}
\left({\frac{\alpha \rho {c_s}^2 R^3 H}{{\Omega}_K}}{\frac{d \Omega}
{dR}}\right)
\end{equation}

\end{itemize}
In the above, $v$ is the radial velocity, $R$ is the radial distance,
$\Omega$ is the angular velocity, ${\Omega}_K$ is the Keplarian 
angular velocity defined as ${{\Omega}_K}^2=GM/R^3$, $c_s$ is the
velocity of sound defined as ${c_s}^2={\partial}P/{\partial}{\rho}$
and $\alpha$ is the effective viscosity of Shakura and Sunyaev~\cite{ss}. 

It is a standard practice to make use 
of a general polytropic equation of state 
$P=k{\rho}^{\gamma}$ where $k$ and $\gamma$ are constants, with $\gamma$ 
being the polytropic exponent~\cite{sc}, whose admissible range 
$(1< \gamma < 5/3)$ is restricted by the isothermal limit and the 
adiabatic limit respectively. The condition of hydrostatic
equilibrium along a direction perpendicular to the thin 
disc, allows for the use of the approximation $(H/R){\cong}
({c_s}/{v_K})$, where ${v_K}=R{\Omega}_K$ ~\cite{fkr}. In that case 
Eq.~(1) leads to 
\begin{equation}
{c_s}^{2n+1}vR^{5/2}={\rm constant}
\end{equation}
where $n=(\gamma -1)^{-1}$. Combining Eqs.~(2) and ~(4) yields,
\begin{equation}
{\frac{d}{dR}}(v^2)={\frac{2v^2}{R}}\left[{\frac{5{c_s}^2/(\gamma +1)-
({{\Omega}_K}^2-{\Omega}^2)R^2}{v^2-2{c_s}^2/(\gamma +1)}}\right]
\end{equation}
and a first integral of Eq.~(3) can be written as 
\begin{equation}
{\alpha} \frac{d}{dR}({\Omega}^2)=2{\Omega}
\left(\Omega -{\frac{L}{R^2}}\right){\frac {v{\Omega}_K}{{c_s}^2}}
\end{equation}
where $L$ is a constant of integration.

We now consider the inviscid limit of Eq.~(6) by setting
$\alpha =0$. This, somewhat simplifying, prescription has found 
regular favour in accretion literature~\cite{az,msc,skc2}, and it gives
the condition ${\Omega}R^2=L$, where $L$, which can be physically
identified as the specific angular momentum of the flow, now becomes 
a constant of the motion. Such a constraint allows us to fix the 
critical points of the flow. At such points both the numerator and
the denominator in the right hand side of Eq.~(5) vanish 
simultaneously~\cite{skc,skc2}, and this will deliver the critical
point conditions as 
\begin{eqnarray}
{v_c}^2 &=& {\frac{2{c_{sc}}^2}{\gamma +1}} \nonumber \\
{\frac{5{c_{sc}}^2}{\gamma +1}} &=& ({{\Omega}_{Kc}}^2-{{\Omega}_c}^2)
{R_c}^2
\end{eqnarray}
with the subscripted label $c$ indicating critical
(fixed) point values. We should be able to find $v_c$ and
$R_c$ from Eqs.~(7). The second relation above, which is actually a
quadratic in $R_c$, leads us to 
\begin{equation}
R_c=\frac{\gamma +1}{10}\frac{GM}{{c_{{sc}}}^2}
\left [ 1 \pm \sqrt{1 - \frac{20}{\gamma +1}
\left ( \frac{c_{{sc}}L}{GM}\right )^2} \right ]
\end{equation}
and this gives the two critical points 
at radii $R=R_{c1}$ and $R=R_{c2}$, with ${R_{c2}}>{R_{c1}}$. To fix the
critical point coordinates $(R_c,v_c)$, we need to look at the integral 
of Eq.~(2), which reads as
\begin{equation}
{\frac{v^2}{2}}+{\frac{L^2}{2R^2}}-{\frac{GM}{R}}+{\frac{{c_s}^2}
{\gamma -1}}={\rm constant}={\frac{{{c_s}^2}(\infty)}{\gamma -1}}
\end{equation}
in which the constant of integration has been determined with the
help of the boundary condition that for very great radial distances, 
the speed of sound approaches a constant ``ambient" value $c_s(\infty)$,
while $v\longrightarrow 0$. In this expression we use the condition 
${v_c}^2=2{{c_{sc}}^2}/(\gamma +1)$ and the value of $R_c$ from 
Eq.~(8), to fix $c_{sc}$ in terms of the constants of the system. 
This will in turn fix the critical point coordinates $(R_c,v_c)$.  

\section{The disc as a dynamical system}

\begin{figure}[t]
\begin{center}
\includegraphics[scale=0.4, angle=-90]{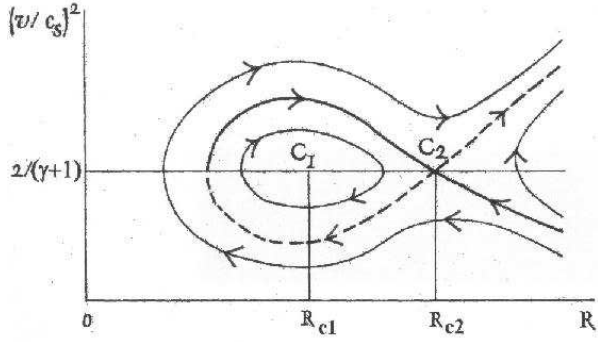}
\caption{\small{A schematic diagram of the accretion problem. The
fixed point $\rm {C_2}$ is a saddle point. $\rm {C_1}$ is a centre
type point. The separatrices pass through $\rm {C_2}$.}}
\end{center}
\end{figure}

To have any understanding of the nature of the critical points of the 
inviscid and thin disc being studied here, we need to cast the equation
giving the stationary solutions of the disc system, along the lines of 
a dynamical system. To do that we appeal to Eq.~(5) and parametrize it 
in the form
\begin{eqnarray}
{\frac{d\left(v^2 \right)}{d \tau}} &=& 2v^2
\left[{\frac{5{c_s}^2}{\gamma +1}}-
\left({{\Omega}_K}^2-{\Omega}^2 \right)R^2 \right] \nonumber \\
{\frac{dR}{d \tau}} &=& R\left[v^2-{\frac{2{c_s}^2}{\gamma +1}} \right]
\end{eqnarray}
It is to be noted that the parametrization has been carried out in an
arbitrary mathematical parameter space $\tau$, and in such a space the two
equations above represent a dynamical system. To analyse the nature of 
the fixed points, we expand and linearize about the fixed point coordinates
$({v_c}^2,R_c)$ in Eqs.~(10), and in terms of the perturbed quantities we 
obtain a set of linear equations given by 
\begin{eqnarray}
{\frac{d}{d \tau}}\left({\delta}v^2\right)
&=& 2{v_c}^2\left[-{\frac{5}{2}}{\frac{\gamma -1}
{\gamma +1}}{\delta}v^2+\left({\frac{6-4{\gamma}}{\gamma +1}}{\frac{GM}
{{R_c}^2}}-{\frac{L^2}{{R_c}^3}}{\frac{7-3{\gamma}}{\gamma +1}}\right)
{\delta}R \right] \nonumber \\
{\frac{d}{d \tau}}\left({\delta}R \right)
&=& {R_c}\left[{\frac{2 \gamma}{\gamma +1}}
{\delta}v^2+5{\frac{\gamma -1}{\gamma +1}}{v_c}^2{\delta}R \right]
\end{eqnarray}
Use of solutions of the form $e^{\lambda \tau}$ would deliver the 
eigenvalues $\lambda$ --- growth rates of ${\delta}v^2$ and 
${\delta}R$ --- as
\begin{equation}
{\lambda}^2=4{\frac{\left(5- \gamma \right)}{\left(\gamma +1 \right)^2}}
\frac{GM}{R_c}{c_{sc}}^2
\left[\beta -{\frac{L^2}{{L_{Kc}}^2}} \right]
\end{equation}
where $\beta = (5-3{\gamma})/(5-{\gamma})$ and the local
Keplerian angular momentum is $L_{Kc} = \sqrt{GMR_c}$.
It is now quite evident that if $R_c$ is the outer fixed point $R_{c2}$,
then ${\lambda}^2$ is positive and the fixed point is saddle type, while
if $R_c$ is the inner point $R_{c1}$, then ${\lambda}^2$ is negative and
the fixed point is centre type. We show the trajectories in Fig.1, in
which the arrows indicate the direction of flows. 

The fixed point $\rm {C_2}$ at $R=R_{c2}$ is the saddle. In a manner of
speaking, it may be dubbed the ``sonic point" because one of the two 
trajectories passing through it is 
transonic, rising from subsonic values far from it to supersonic 
values when $R<R_{c2}$. This trajectory is shown as the heavy solid curve,
and is the accretion flow. The heavy dashed curve is the wind.  The 
above demonstration that the fixed point $\rm {C_2}$ is a saddle has adverse 
implications regarding the realizability of the transonic trajectory
passing through it. Conventional wisdom about the nature of saddle type
points~\cite{js} gives us to believe that if 
we go directly to the static limit of the dynamic evolution, then the
transonic trajectory would not be realizable. However, it is 
generally believed that transonic trajectories do occur in 
astrophysical systems such as the one being studied here~\cite{skc}. 
Our contention is that it is the temporal evolution of such a system 
that selects the transonic trajectory. This selection mechanism is purely 
non-perturbative as opposed to the perturbative technique of a linear
stability analysis of all physically feasible stationary inflow solutions.  
The latter method offers us no clue as to the choice of the system for
any particular solution, since all stationary solutions have been found 
to be stable under the influence of a linearized 
perturbation --- considered
both as a travelling wave and a standing wave~\cite{ray}.

As an aside it may also be mentioned here that the nature of the outer
fixed point $\rm {C_2}$ in Fig.1, has a bearing 
on the possible range of values that the constant specific angular 
momentum $L$ may be allowed. For solutions passing through the outer
critical point, it can be well recognized that the flow 
would be sub-Keplerian~\cite{skc,az}. In that case, it may be concluded 
that the condition $0<(L/L_{Kc2})^2<1$ would hold good. In addition to 
that, from Eq.~(12), the saddle type behaviour of the point $\rm {C_2}$ 
would also imply that there would be another upper bound on $L$, given 
by $(L/L_{Kc2})^2< \beta$. For the admissible range of
the polytropic index $\gamma$, the possible range for $\beta$ 
would be $0< \beta < 1/2$. This would then imply that the latter bound
on $L$ would be more restrictive, as compared to the former. It is 
interesting that this essentially physical conclusion could be drawn
from modelling the disc as a dynamical system in a mathematical parameter
space. 

\section{A model for a dynamic selection mechanism}

In the previous section we contended that the temporal evolution of the
disc system selects a critical solution that passes through the outer
critical point. To understand how this comes about, we consider
a simplified mathematical example, since it is well known that the 
equations of a compressible flow cannot be exactly integrated. 

The model system that we introduce, describes the dynamics of $y(x,t)$ as
\begin{equation}
{\frac{\partial y}{\partial t}}+\left(y-x \right)
{\frac{\partial y}{\partial x}} =y+2x+x^2
\end{equation}
whose static limit yields 
\begin{equation}
{\frac{dy}{dx}}={\frac{y+2x+x^2}{y-x}}
\end{equation}
and which, viewed as dynamical system, is seen as 
\begin{eqnarray}
{\frac{dy}{d \tau}} &=& y+2x+x^2 \nonumber \\
{\frac{dx}{d \tau}} &=& y-x
\end{eqnarray}

In the $y-x$ space, the fixed points $(x_c,y_c)$ are to be found at
$(0,0)$ and $(-3,-3)$. As in the preceding section, a linear stability
analysis of the fixed points in $\tau$ space gives the eigenvalues 
$\lambda$ by ${\lambda}^2 = 1+2(x_c +1)$.  It is then easy to see that 
$(0,0)$ is a saddle point while $(-3,-3)$ is a centre type point. 
The integral curves are 
\begin{equation}
y^2-2xy-2x^2-{\frac{2}{3}}x^3=c
\end{equation}
and 
the trajectories passing through the saddle point are the ones with
$c=0$. The different possible trajectories are shown in Fig.2.
The separatrices are shown as the heavy solid curve and the heavy dashed 
curve. The similarity between Figs.1 \& 2 is obvious. 
We now explore the dynamics. 

To obtain a solution to Eq.~(13), we need to apply the method
of characteristics~\cite{deb}. This involves writing
\begin{equation}
{\frac{dt}{1}}={\frac{dx}{y-x}}={\frac{dy}{y+2x+x^2}}
\end{equation}
The task is to find two constants $c_1$ and $c_2$ from the above set
and the general solution of Eq.~(13) would be given by ${c_1}
=F(c_2)$, where the function $F$ is to be determined from the initial
conditions. It is easy to see that one of the constants of integration
is clearly the $c$ of Eq.~(16). Hence, we write ${c_1}=c$ and
from the first part of Eq.~(17), we can write
\begin{equation}
\int dt={\pm}{\int}{\frac{dx}{\sqrt{3x^2+\left(2/3 \right)x^3+c}}}
\end{equation}
which solves the problem in principle. To put this in a usable form,
we need to carry out the integration of Eq.~(18). This cannot
be done exactly. For small $x$ (the most important region, since it 
is near the saddle), we can drop the $x^3$ term to a good approximation. 
Further, we choose only the positive sign in the right hand side of
Eq.~(18) by the physical argument that we would wish to evolve the 
system through a positive range of $t$ (time) values. Integration of
Eq.~(18) will then lead us to the result
\begin{equation} 
\left(x+ \sqrt{x^2+c/3} \right)e ^{- \sqrt{3} t}={c_2}
\end{equation}
which will then make the solution of Eq.~(13) look like
\begin{equation}
y^2-2xy-2x^2-{\frac{2}{3}}x^3=F \left(\left[x
+ \sqrt{{\frac{\left(y-x \right)^2}{3}}-
\frac{2}{9}x^3}\right]e^{-{\sqrt 3}t}\right) 
\end{equation}

\begin{figure}[t]
\begin{center}
\includegraphics[scale=0.4, angle=-1.0]{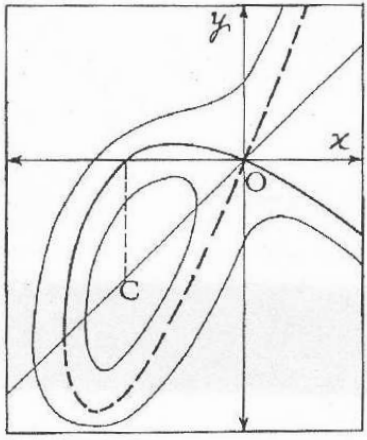}
\caption{\small{A schematic diagram of the model problem. The curve
given by Eq.~(16) passes through the origin for $c=0$. For
this critical trajectory $\rm {O}$ is a saddle point while $\rm {C}$
is a centre type fixed point. This figure differs from Fig.1 by a
tilt of the axes.}}
\end{center}
\end{figure}

We wish to study the evolution of the system from the initial 
condition $y=0$ for all $x$ at $t=0$. For small $x$, dropping the
$x^3$ term again, allows us to determine the form of the function 
$F$ as $F(z)=-3(2- \sqrt{3})z^2$. The solution, consequently, becomes 
\begin{equation}
y^2-2xy-2x^2-{\frac{2}{3}}x^3=-3\left(2- \sqrt{3}\right)
\left[x+ \sqrt{{\frac{\left(y-x \right)^2}{3}}-
\frac{2}{9}x^3}\right]^2
e^{-2{\sqrt{3}}t}
\end{equation}
and as $t{\longrightarrow}{\infty}$, we select the separatrices
given by 
\begin{equation}
y^2-2xy-2x^2-{\frac{2}{3}}x^3=0
\end{equation}
a result whose remarkable feature is worth stressing. The
evolution started far away from the separatrix. In fact, it started
as $y=0$ for all $x$ at $t=0$. The evolution proceeded through a 
myriad of stationary states (all arguably stable under a linear stability 
analysis) and then locked on to the separatrix. Our claim is that
this dynamic mechanism selects the velocity profile in the practical
situation as well, and this is entirely in consonance with a similar
and well established selection mechanism for the spherically symmetric
flow~\cite{rb}. 

\section{Concluding remarks}

In a previous section we discussed that linear stability analysis
of the stationary inflow solutions offers no direct clue as to the 
preference of the disc system for a particular stationary solution. 
However, in a somewhat indirect approach to this issue we could make 
use of a linear stability analysis to have a feel for  
the special feature of the critical solution. A very convenient way of
carrying out the linear stability analysis is in terms of the variable
${\phi}={\rho}^{(\gamma +1)/2}vR^{5/2}$~\cite{ray}. In terms of this 
variable, the perturbation equation for $\phi$ reads
\begin{equation}
{\frac{{\partial}^2 {\phi}^{\prime}}{\partial t^2}}+2{\frac{\partial}
{\partial R}}\left({v_0}{\frac{\partial {\phi}^{\prime}}{\partial t}}\right)
+ {\frac{1}{v_0}}{\frac{\partial}{\partial R}}\left[{v_0}\left({v_0}^2
-{\frac{2}{\gamma +1}}{c_{s0}}^2\right)
{\frac{\partial {\phi}^{\prime}}{\partial R}}\right]=0
\end{equation}
in which ${\phi}^{\prime}$ is the perturbation on the steady solution of
$\phi$, and the subscripted label $0$ indicates steady state 
values wherever in use~\cite{ray}.

In a spacetime with metric $g^{\mu \nu}$, the wave equation for a 
scalar variable is~\cite{vis} 
\begin{equation}
\frac{\partial}{\partial x^\mu}\left(g^{\mu \nu}
\frac{\partial \phi^\prime}{\partial x^\nu}\right)=0
\end{equation}
Ignoring the angle variables, and using the notation that ${\mu}=1$
corresponds to $t$ and ${\mu}=2$ corresponds to $R$, we have the 
identification
\begin{equation}
g^{11}=1, \,\, g^{12}=g^{21}={v_0}, \,\, g^{22}={v_0}^2-{\frac{2}{\gamma +1}}
{c_{s0}}^2
\end{equation}
The effective covariant components are 
\begin{equation}
g_{11}=1-{\frac{{v_0}^2}{{\tilde{c_{s0}}}^2}}, \,\, g_{12}=g_{21}={\frac
{v_0}{{\tilde{c_{s0}}}^2}}, \,\, g_{22}=-{\frac{1}{{\tilde{c_{s0}}}^2}}
\end{equation}
in which ${\tilde{c_{s0}}} = \sqrt{2}(\gamma + 1)^{-1/2}{c_{s0}}$. 

The differential distance in the metric space has the form
\begin{equation}
ds^2={\frac{1}{{\tilde{c_{s0}}}^2}}\left[
{{\tilde{c_{s0}}}^2}dt^2-\left(dx^i-{{v_0}^i}dt \right)
{{\delta}_{ij}}\left(dx^j-{{v_0}^j}dt\right) \right]
\end{equation}
and this is to be compared with the Painlev\'e-Gullstrand form of 
$ds^2$ for a rotating black hole of mass $\tilde{M}$
\begin{equation}
ds^2=dt^2-\left(dr{\pm}{\sqrt{{\frac{2G{\tilde{M}}}{r}}
-{\frac{L^2}{r^2}}}}dt \right)^2 +{r^2}d{\theta}^2
\end{equation}

We now note from Eq.~(7) that at the sonic point,
${v_c}^2=(2/5)[(GM/{R_c})-(L/{R_c})^2]$, and hence in the vicinity 
of the sonic point the metric implied by the linear stability analysis
of the stationary solution appears quite close in form to the 
metric of a rotating black hole. The unidirectional nature of
trajectories near a black hole suggests that trajectories go 
through the sonic point, which is equivalent to saying that
transonic trajectories will be chosen.

In concluding we would also like to comment on the nature of the possible
critical point(s) when we consider the more likely case of a disc with 
a mechanism for the
outward transport of angular momentum. It is possible with the help of
Eq.~(6) to expand and linearize about the critical point value
of the angular velocity $\Omega$, but without any knowledge of an
exact dependence of $\Omega$ on $R$, there would be no way of fixing 
the critical point(s). However it can be conceived that as an extension
of the inviscid case, there may be multiple parameter-dependent
critical points~\cite{skc3}
and in that event at least one of them may be saddle type, with all its
associated difficulties within the stationary framework itself. In that
event, once again the selection mechanism has to be time evolutionary
in nature. 

\section*{Acknowledgements}
This research has made use of NASA's Astrophysics Data System. One of
the authors (AKR) would like to acknowledge the financial assistance
given to him by the Council of Scientific and Industrial Research, 
Government of India.

\end{document}